\documentclass{aa}

\usepackage{amsmath}
\usepackage{graphics}

\begin{document}

\thesaurus{02(09.04.1; 11.09.3; 12.04.1)}

\title{Can Dust Segregation Mimic a Cosmological Constant?}

\author{Jakob T.\ Simonsen \and
Steen Hannestad}

\institute{Institute of Physics and Astronomy,
University of Aarhus,
DK-8000 \AA rhus C, Denmark}

\date{\today}

\maketitle

\begin{abstract}
Recent measurements of type Ia supernovae indicate that distant
supernovae are substantially fainter than expected from the
standard flat cold dark matter model. One possible explanation
is that the energy density in our universe is in fact dominated
by a cosmological constant. Another possible solution is that
there are large amounts of grey dust in the intergalactic medium.
Dust grains can be grey either because they are non-spherical
or very large. We have numerically 
investigated whether grey dust can be emitted from high
redshift galaxies without also emitting standard, reddening dust,
which would have been visible in the spectra of high redshift
objects.
Our finding is that grain velocities are almost independent of
ellipticity so that if greyness are due to the grains being
elongated, it will not be possible to separate grey dust from
ordinary dust.
We also find that velocities are fairly independent of grain
size, but we cannot rule out possible sputtering of small grains, so
that large, grey dust grains could be preferentially emitted.
Therefore, our conclusion is that
grey dust is an unlikely explanation of the data, but we cannot
rule it out if the grey dust consists of large, spherical grains.


\end{abstract}

\section{Introduction}

In recent years the prospect of using type Ia supernovae for
distance measurements in the universe has attracted a great
deal of attention. Many studies indicate that these supernovae
are very close to being standard candles, so that measuring their
effective magnitude amounts to directly estimating their luminosity
distance. Two large scale surveys of high redshift supernovae are
currently in progress 
(Perlmutter {et al.} \cite{perlmutter98}, \cite{perlmutter97}; 
Garnavich {et al.} \cite{garnavich};
Riess {et al.} \cite{riess}; Schmidt {et al.} \cite{schmidt}), 
and both groups have published results for the determination of the
deceleration parameter, $q_0$.
Surprisingly, the results are statistically incompatible
with the value $q_0=1/2$ expected in the standard, flat matter
dominated Friedmann model (Kolb \& Turner \cite{kt90}).
The observed high redshift supernovae are significantly fainter
than expected for $q_0=1/2$. Rather they are compatible with a
medium density, flat universe, where roughly 70\% of the
energy density comes from vacuum energy (Perlmutter {et al.} 
\cite{perlmutter98})

If this turns out to be true it is an astonishing result because
it means that, taken together with the recent results on neutrino
masses from Super-Kamiokande (Fukuda {et al.} \cite{fukuda}), 
there are at least four components
contributing almost equally to the cosmic energy density budget:
baryons, neutrinos, cold dark matter and vacuum energy. 
Even without vacuum energy this is a remarkable fine tuning, but 
including vacuum energy as a fourth component makes the conceptual
problem substantially worse.

For this reason it is important to check all alternative explanations
of the supernova data that can be thought of. One such possibility
is evolution effects in the supernova progenitors. However, the data
seem to indicate that the supernova population at high redshift
is remarkably similar to the one known at low redshift
(Perlmutter {et al.} \cite{perlmutter98};
H{\"o}flich, Wheeler \& Thielemann \cite{hwt98}).
Another intriguing possibility which has been advocated by Aguirre
(Aguirre \cite{aguirre}, \cite{aguirre2})
is that there could be unseen dust contamination in the data.
In the data reduction it has been assumed that any possible dust
follows the standard extinction law known to govern dust in the
Milky Way (Cardelli, Clayton \& Mathis \cite{ccm89}).

Based on this, almost no dust is found in the line of sight
to any of the supernovae. However, this conclusion is based on 
the fact that dust in the Milky Way
reddens significantly, simply because
it is primarily made up of small, spherical grains
(Draine \& Lee \cite{drlee}; Aguirre \cite{aguirre2}).
This need not be the case. If there is some population of very large
and/or elongated
dust grains which is homogeneously distributed in the universe,
 this would remain undetected
in reddening surveys because of its very flat extinction curve
(Aguirre \cite{aguirre}, \cite{aguirre2}), i.e.\ it would be ``grey'' dust. 
In fact the idea of grey dust is quite old
and has been advocated as a means of thermalising the cosmic 
microwave background radiation in the steady state cosmology
(see for instance Peebles \cite{peebles} for a review).

The above scenario could perhaps be a viable explanation of the data, 
given that
enough such dust can be expelled from galaxies at high redshift
{\it without} simultaneously expelling small dust grains.

The purpose of the present paper is to quantify this by numerical
modelling of dust expulsion. We have wished to investigate whether
it is in fact possible to expel only very elliptical and large grains.
We have not worried about the possible production mechanisms
for dust grains, only about their dynamical
behaviour. A completely different problem is whether it is at
all possible to produce the required amounts of large dust grains
in the first place.
However, by using our relatively simple model we are able to make the
very robust prediction that dust segregation at the level needed
to explain the data is not possible, unless very special conditions
prevail in the host galaxies.
Note that in the present paper we only consider radiation pressure
as a source of dust expulsion. There could be several other ways 
of removing dust from galaxies, such as winds from supernovae
(Suchkov {et al.} \cite{suchkov}) or galaxy collisions (see e.g. Gnedin
\cite{gnedin}). However, these methods seem unlikely to preferentially emit
large dust grains, and for that reason we have, in accordance with
Aguirre (\cite{aguirre}, \cite{aguirre2}) 
concentrated on radiation pressure, even though
galaxy collisions might be the most efficient method of dust
removal at high redshift (Gnedin \cite{gnedin}).

The paper has been sectioned as follows: Section 2 contains a
discussion of the equation of motion for dust grains in a galactic
environment, section 3 contains our main numerical results, and
finally section 4 is devoted to a discussion of our findings.

\section{Equation of motion for dust grains in a galactic medium}

At present the supernovae with the highest measured redshifts are
at $z \simeq 0.8$ (Perlmutter {et al.} \cite{perlmutter98}). 
This means that if dust is to explain the data
it must be expelled and distributed uniformly prior to this epoch.
For this to happen it must have been expelled from the host galaxies
at very high redshift, probably $z \simeq 3$ or more
(Aguirre \cite{aguirre}, \cite{aguirre2}).
Relatively little is known about the structure of galaxies at such 
distances, except that they are richer in gas content than our own
galaxy at present.

We have taken a heuristic approach to our lack of knowledge about
the nature of the host galaxies and calculated dust expulsion for
a range of different gas contents and emission temperatures.
We find that our results depend very little on these assumptions.

We assume that a given spiral galaxy can be reasonably
well approximated
by three components: disk, bulge and halo. The disk 
is taken to be an infinitely thin axially symmetric exponential 
distribution of matter. The bulge, which is massive and luminous, is 
just added to the center of the galactic disk, and the halo is a 
spherical symmetric mass structure considered to be completely dark.

The three main contributions to the force acting on a dust grain
 moving through a galaxy are gravity, radiative forces and drag forces. 
As the galaxy is assumed to be axially symmetric, all calculations are
 done in cylindrical coordinates $(R,z,\phi)$.

In the following we shall always work with ellipsoidal dust grains, with
semiaxes $a$ and $b$. The effective radius of such a grain is then defined
by the relation
\begin{equation}
V \equiv \frac{4}{3} \pi a_{\rm eff}^3 = \frac{4}{3} \pi a^2 b.
\end{equation}

\subsection{Gravitational forces}

The gravitational force on a dust grain of mass $m_\mathrm{g} \equiv 
\rho_g V$, where $\rho_g$ is the grain density, 
is simply given by
\begin{equation}
\mathbf{F}_\mathrm{G} = m_\mathrm{g} \mathbf{G}(\mathbf{r}),
\end{equation}
where $\mathbf{G}(\mathbf{r})$ is the gravitational field intensity 
at the point $\mathbf{r}$.

The disk has an exponential mass distribution with surface density
\begin{equation}
\Sigma(R) = \Sigma_0 \exp(-R/R_\mathrm{d}),
\end{equation}
where $R_\mathrm{d}$ is the disk scale length. 
The potential generated by this disk is most easily expressed via 
Bessel functions, and takes the form (see e.g.\ Binney \& Tremaine 
\cite{binney})
\begin{equation}
\Phi(R,z) = -2\pi G\Sigma_0 R_\mathrm{d}^{2}\int_{0}^{\infty}
\frac{J_0(kR)\exp(-kz)}{[1+(kR_\mathrm{d})^2]^{3/2}}\mathrm{d}k,
\end{equation}
$J_n(x)$ being the Bessel function of the first kind of order n.
The gravitational force from the disk on a dust grain with mass 
$m_\mathrm{g}$ is thus given by
\begin{equation}\begin{align}
F_{\mathrm{grav,disk},R}& = -m_\mathrm{g} 
2\pi G\Sigma_0 R_\mathrm{d}^{2}\int_{0}^{\infty}
\frac{kJ_1(kR)\exp(-kz)}{[1+(kR_\mathrm{d})^2]^{3/2}}\mathrm{d}k, \nonumber \\
F_{\mathrm{grav,disk},z}& = -m_\mathrm{g} 
2\pi G\Sigma_0 R_\mathrm{d}^{2}\int_{0}^{\infty}
\frac{kJ_0(kR)\exp(-kz)}{[1+(kR_\mathrm{d})^2]^{3/2}}\mathrm{d}k.
\label{eq:dgrav}
\end{align}\end{equation}

The potential from the bulge is that of a point mass with mass $M_\mathrm{b}$. 
That is,
\begin{equation}
F_{\mathrm{grav,bulge},r} = -\frac{GM_\mathrm{b} m_\mathrm{g}}{r^2},
\end{equation}
where $r = \sqrt{R^2 + z^2}$. Projecting on to the $R$ and $z$ axis gives
\begin{equation}\begin{align}
F_{\mathrm{grav,bulge},R}& = -\frac{GM_\mathrm{b} m_\mathrm{g}}
{r^2}\frac{R}{r} = -G M_\mathrm{b} m_\mathrm{g} \frac{R}{(R^2+z^2)^{3/2}}.
\nonumber \\
F_{\mathrm{grav,bulge},z}& = -\frac{GM_\mathrm{b} m_\mathrm{g}}
{r^2}\frac{z}{r} = -G M_\mathrm{b} m_\mathrm{g} \frac{z}{(R^2+z^2)^{3/2}}.
\label{eq:bulgrav}
\end{align}
\end{equation}

For the halo we assume an isothermal sphere distribution
\begin{equation}
\rho = \rho_0 \frac{a_h^2}{a_h^2 + r^2}.
\end{equation}
Integrating the equation for $\rho$ yields
\begin{equation}
M(r) = 4\pi \rho_0 a_\mathrm{h}^{2}[r - a_\mathrm{h} \arctan(r/a_\mathrm{h})],
\end{equation}
where $\rho_0$ is the central density and $a_\mathrm{h}$ is the halo 
mass scale length. Setting $M_\mathrm{h} = M(r)$, this results in a 
central force, which after projecting on to the $R$ and $z$ axis takes the form
\begin{equation}\begin{align}
F_{\mathrm{grav,halo},R}& = -G M_\mathrm{h} m_\mathrm{g} \frac{R}{(R^2+z^2)^{3/2}},
\nonumber \\
F_{\mathrm{grav,halo},z}& = -G M_\mathrm{h} m_\mathrm{g} 
\frac{z}{(R^2+z^2)^{3/2}}. \end{align}
\end{equation}

\subsection{Radiative forces}

The radiation pressure force can be written in a way completely analogous 
to the gravitational force (Ferrara {et al.} \cite{ferrara}).

First, we shall assume that the spectral function, $\Omega_\nu$,
of the emitted light is constant
throughout the galaxy, i.e.\
\begin{equation}
\Omega_\nu (R,z) = \Omega_\nu.
\end{equation}
If this is the case then we can calculate a spectrally averaged radiation
pressure coefficient for a given dust grain as
\begin{equation}
Q_{\rm pr}^* = \int d \nu \Omega_\nu Q_{\rm pr} (\nu),
\end{equation}
which applies at all positions in the galaxy. 
In practise we shall assume that the galaxy emits light as a black body with
some temperature $T_0$, so that $\Omega_\nu = B_\nu (T_0)$, where $B_\nu$ 
is the Planck function, and $Q_\mathrm{pr}^*(a_\mathrm{eff}) = 
Q_\mathrm{pr}(a_\mathrm{eff},T_0)$.

The disk luminosity distribution is assumed exponential like the mass distribution, 
and with the same scale length $R_\mathrm{d}$
\begin{equation}
I(R) = I_0 \exp(-R/R_\mathrm{d}).
\end{equation}
From these simple assumptions the radiation pressure force from the disk 
is given as
\begin{equation}\begin{align}
F_{\mathrm{rad,disk},R}& = \frac{2\pi^2 a_\mathrm{eff}^{2} 
Q^*_\mathrm{pr}I_0}{c}\int_{0}^{\infty}\frac{kJ_1(kR)\exp(-kz)}
{[1+(kR_\mathrm{d})^2]^{3/2}}\mathrm{d}k, \nonumber  \\
F_{\mathrm{rad,disk},z}& = \frac{2\pi^2 a_\mathrm{eff}^{2} 
Q^*_\mathrm{pr}I_0}{c}\int_{0}^{\infty}\frac{kJ_0(kR)\exp(-kz)}
{[1+(kR_\mathrm{d})^2]^{3/2}}\mathrm{d}k,
\end{align}\end{equation}
analogous of Eq.~(\ref{eq:dgrav}).

The force from the bulge is similarly given as 
\begin{equation}
F_{\mathrm{rad,bulge},r} = \frac{\pi a_\mathrm{eff}^{2} 
Q^*_\mathrm{pr}}{c}\frac{L_b}{4\pi r^2},
\end{equation}
where $L_b$ is the total luminosity of the bulge.
In the $R$ and $z$ directions this is
\begin{equation}\begin{align}
F_{\mathrm{rad,bulge},R}& = \frac{a_\mathrm{eff}^{2} 
Q^*_\mathrm{pr}L_\mathrm{b}}{4c}\frac{R}{(R^2+z^2)^{3/2}}, \nonumber \\
F_{\mathrm{rad,bulge},z}& = \frac{a_\mathrm{eff}^{2} 
Q^*_\mathrm{pr}L_\mathrm{b}}{4c}\frac{z}{(R^2+z^2)^{3/2}},
\end{align}\end{equation}
again, apart from numerical factors equal to Eq.~(\ref{eq:bulgrav}).

Note that there is an important issue which we do not address here, namely the 
opacity in the disk itself (Davies {et al.} \cite{davies}). 
Since the disk contains significant amounts of dust
it will obscure the starlight and lead to smaller radiation pressure. This was
determined to be an important effect for the dynamical evolution of grains
by Davies {et al.} (\cite{davies}). 

However, it is not entirely clear how disk opacity affects the grain
evolution. It leads to smaller radiation pressure, but on the other hand it
also reddens the light, leading to a lower effective temperature of the
radiation field.
We model this behaviour by calculating grain evolution for several different
emission temperatures.

\subsection{Rotation}

The stars and gas in the disk of spiral galaxies have purely rotational orbits.
This means that the rotational velocity is given by 
\begin{equation}
v_{\rm rot} = \sqrt{R \frac{\partial \Phi}{\partial R}}.
\end{equation}
Initially the grain also rotates with this velocity, but as the grain moves outwards
the rotational velocity decreases. If gas drag is neglected the angular momentum
in the $z$ direction is conserved so that 
\begin{equation}
v_{\rm rot}(R) = v_{\rm rot}(R_i) (R_i/R).
\end{equation}
However, for typical spiral galaxies the rotational velocity of the stars
remains constant out to very large values of $R$. This means that grains
will, in general, be spun up by interaction with the disk gas. This effect
is in practice quite small, but we have included it anyway.

\subsection{Drag forces}

The interaction of dust grains with gas plays an important role in grain 
evolution. Assuming that the grain moves fast compared with the thermal 
velocities in the gas, the exact expression for the drag force experienced 
by a grain is (Il'in \cite{ilin})
\begin{equation}
{\bf F}_\mathrm{drag} =- \pi ab\rho_\mathrm{gas} v_\mathrm{g}^{2} 
\frac{2}{\pi} E([1 - e^2] \sin^2\Theta) \hat{\bf v}_\mathrm{g}
\label{eq:drag}
\end{equation}
where $a$ and $b$ are the semiaxes of the spheroidal grain, $e=a/b$, 
${\bf v}_\mathrm{g}$ the grain velocity relative to the gas, 
$E(x)$ the complete elliptic 
integral of the second kind and $\Theta$ the angle between magnetic 
field direction and wave vector of radiation. 

Interstellar dust grains rotate with large angular velocity because of
the angular momentum transferred to them by collisions. 
Non-spherical grains are likely to rotate along the axis of maximum
moment of inertia, with the angular momentum being aligned along the
magnetic field lines (Il'in \cite{ilin}).
For this reason $\Theta$ is likely to be small 
so that the argument to the elliptic integral 
is close to 0. Since $E(0)=\pi/2$ the drag force becomes
\begin{equation}
{\bf F}_\mathrm{drag} = -\pi ab\rho_\mathrm{gas} v_\mathrm{g}^{2} 
\hat{\bf v}_\mathrm{g}.
\end{equation}
As mentioned above, this drag force also causes the grain to be spun up 
by the gas, so that, as long as it stays in the gas, ${\bf v}_\phi$, is driven
towards the rotational velocity of the gas.
Note that $[1-e^2] \sin^2 \Theta \in [0,1]$ so that
$E(x) \in [1,\pi/2]$ always, i.e.\ choosing $x=0$ produces at worst an error
of 50\% in the drag force.

Also, the above formula only applies if the grain velocity is larger than
the thermal velocity of the gas. However, using Eq.(\ref{eq:drag}) leads to
small errors, as will be discussed in section 3.1.

\subsection{Sputtering}

In general, grains are eroded as they pass through the galactic medium, by collisions
with gas particles and by radiation (Draine \& Salpeter \cite{ds79}, 
Ferrara {et al.} \cite{ferrara}, Shustov \& Vibe \cite{shustov}). 
The amount of sputtering increases
with gas density, both because of the higher collision rate and because the grain
spends more time in the region with high gas density. 
The effect of sputtering is primarily to destroy
very small dust grains, but leave large dust grains more or less unscathed
(Ferrara {et al.} \cite{ferrara}, Davies {et al.} \cite{davies}).
It is very difficult to assess the amount of sputtering in a given galaxy.
For galaxies similar to the Milky Way, Ferrara et al.\ (1991) find that
sputtering is not important unless the grains are very small ($a \sim 10-20$ nm),
whereas Shustov \& Vibe (1995) find that grains up to 50 nm could be destroyed.
We can therefore not rule out possible destruction of grains up to about this size.

\subsection{Magnetic forces}

As has been noted many times in the literature, galactic magnetic fields are
important for the dynamical evolution of dust. However, very little is known 
about the nature of these fields, so we have chosen to neglect possible 
magnetic fields in the present study. In general, the effect of the magnetic 
field will be to enhance the confinement of grains because the drag force 
on grains is enhanced (Ferrara {et al.} \cite{ferrara}) and because 
they are forced to move along the field lines (Ferrara {et al.} 
\cite{ferrara}; Davies {et al.} \cite{davies}).

\section{Numerical results}

In order to do actual numerical calculations it is necessary to specify relevant
parameters for the host galaxy. In practice we have taken the present day Milky
Way with parameters as specified in Table 1 
(Binney \& Tremaine \cite{binney}, Ferrara {et al.} \cite{ferrara}). 
This may not be a very accurate 
description of a typical high redshift galaxy, but as will be shown the numerical
results are fairly insensitive to these assumptions. The most important
issues are the emission temperature of the galaxy and the amount of gas 
contributing to the drag force. We have modelled the radiation field of the 
galaxy as a black body with temperature, 
$T_0 = 6000$ K, 9000 K or 12000 K.

\begin{table}\centering
\caption{Physical parameters for our sample galaxy (the Milky Way).}
\begin{tabular}{llc}\hline\hline
&Parameter & value \\ \hline
Disk&$\Sigma_0$ & 381.9 M$_{\odot}$ pc$^{-2}$ \\
&$R_d$ & 5 kpc \\
&$I_0$ & 108.28 L$_{\odot}$ pc$^{-2}$ \\
Bulge&$M_b$ & $1.0 \times 10^{10}$ M$_{\odot}$ \\
&$L_b$ & $1.5 \times 10^9$ L$_{\odot}$ \\ 
Halo&$\rho_0$ & $6.3 \times 10^{-24}$ g cm$^{-3}$ \\
&$a_h$ & 2.8 kpc \\ \hline
\end{tabular}
\end{table}

\begin{figure}[t]
\resizebox{\hsize}{!}{\includegraphics{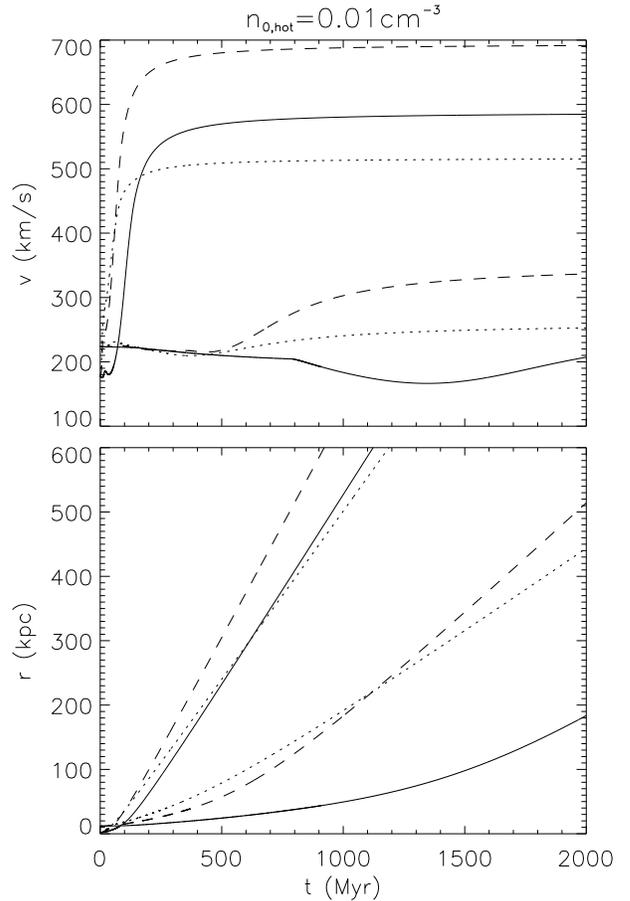}}
\caption{Grain velocity $v$ and displacement $r$ for the present day Milky
Way model, with 
$(a_\mathrm{eff},e)=(50\thinspace\mathrm{nm},1), 
(50\thinspace\mathrm{nm},1/8),(300\thinspace\mathrm{nm},1)$ (full, dashed
and dotted curves respectively), and initial galactocentric distance
$R_0=2\thinspace\mathrm{kpc}$ (upper three curves) and
$R_0=12\thinspace\mathrm{kpc}$ (lower three curves). Notice that the initial
velocity is due to rotation.}
\label{tvr01}
\end{figure}

\begin{figure}[t]
\resizebox{\hsize}{!}{\includegraphics{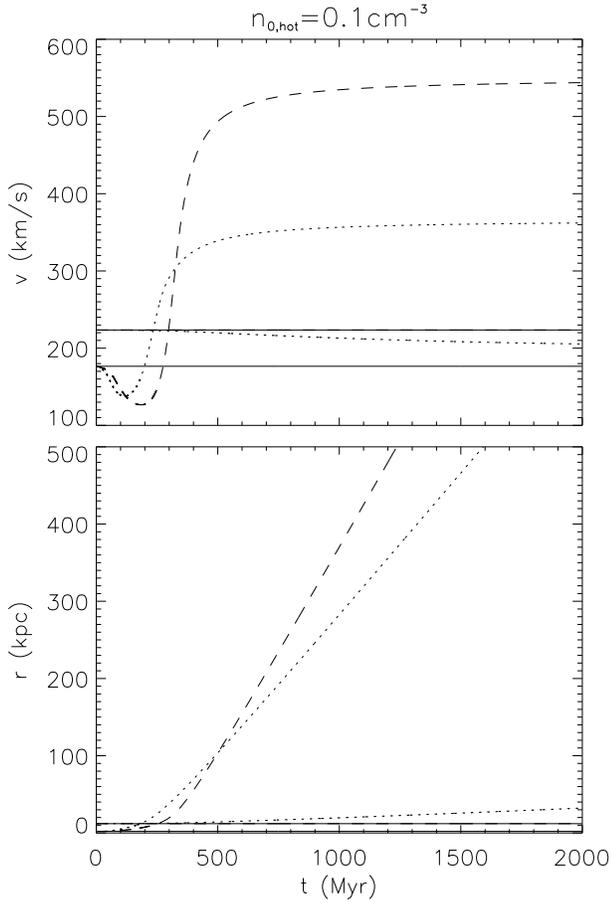}}
\caption{Grain velocity $v$ and displacement $r$ for our extremely
gas rich model, with 
$(a_\mathrm{eff},e)=(50\thinspace\mathrm{nm},1), 
(50\thinspace\mathrm{nm},1/8),(300\thinspace\mathrm{nm},1)$ (full, dashed
and dotted curves respectively), and initial galactocentric distance
$R_0=2\thinspace\mathrm{kpc}$ (lowest initial v) and
$R_0=12\thinspace\mathrm{kpc}$ (highest initial v). Notice that two of the
$R_0=12\thinspace\mathrm{kpc}$ curves overlap, and that the initial velocity
is due to rotation.}
\label{tvr1}
\end{figure}

Furthermore we have considered three different models for the gas distribution.
In all three cases we model the gas by two different components, a cold gas
and a hot halo gas. Both components are assumed to have the 
number density distribution
\begin{equation}
n(R,z) = n_0 e^{-z/z_s},
\end{equation}
i.e.\ constant density as a function of $R$ and an exponential fall-off with 
$z$. The cold component is assumed to have $z_s = 170$ pc, and the hot 
$z_s = 2$ kpc.
For the three different models we then use the following values for $n_0$\\
1: \, $n_{0,{\rm hot}}=1.0 \times 10^{-4}\, {\rm cm}^{-3}$, 
$n_{0,{\rm cold}}=2.0 \times 10^{-3}\, {\rm cm}^{-3}$\\
2: \, $n_{0,{\rm hot}}=1.0 \times 10^{-2}\, {\rm cm}^{-3}$, 
$n_{0,{\rm cold}}=2.0 \times 10^{-1}\, {\rm cm}^{-3}$\\
3: \, $n_{0,{\rm hot}}=1.0 \times 10^{-1}\, {\rm cm}^{-3}$, 
$n_{0,{\rm cold}}=2.0 \, {\rm cm}^{-3}$\\
The first model corresponds to the extreme case of an almost empty galaxy, 
the second
roughly to the present day Milky Way (Lockman \cite{lockman}; Heiles 
\cite{heiles}), and the third to an extremely gas enriched
galaxy. In all cases we assume that gas is made entirely of hydrogen.
Note that we have assumed a model with constant density as a function of
$R$. This may seem a strong assumption, but as will be shown in the next section
the numerical results depend relatively little on it.

The next important issue is the physical composition of the dust grains.
Aguirre \cite{aguirre} assumed pure graphite grains, but relaxed 
that assumption in a subsequent study (Aguirre \cite{aguirre2}), where 
a more realistic galactic dust composition of silicate and graphite 
was assumed.

Again, we have taken an heuristic approach and calculated our models for two
different cases, amorphous carbon and graphite-like carbon; silicates are
expected to behave very similarly. 
To find the radiation pressure coefficient for these grains we have used
dielectric data from J\"ager {et al.}\ \cite{jager}. They have obtained 
dielectric data on carbon pyrolised in the laboratory at different
temperatures. Their 400 C sample has an amorphous structure while their
1000 C sample closely resembles graphite. Both are bulk samples and therefore
isotropic. Note that the graphite needles used in the Aguirre (\cite{aguirre})
model are highly anisotropic, i.e. they are macrocrystals, whereas our
data comes from a bulk sample containing microcrystals. However, in practice
the difference between these two assumptions is likely to be small,
so that our results apply to the Aguirre (\cite{aguirre}) scenario.
We take these two samples as extreme cases of carbon composition in dust 
grains. Using this data we have proceeded to calculate the radiation 
pressure coefficient
by use of the Discrete Dipole approximation as implemented in the DDSCAT
package (Draine \& Flatau \cite{draine}). The resulting $Q_\mathrm{pr}^*$ 
are tabulated in \ref{qtables}.

\begin{figure*}[t]
\resizebox{\hsize}{!}{\includegraphics{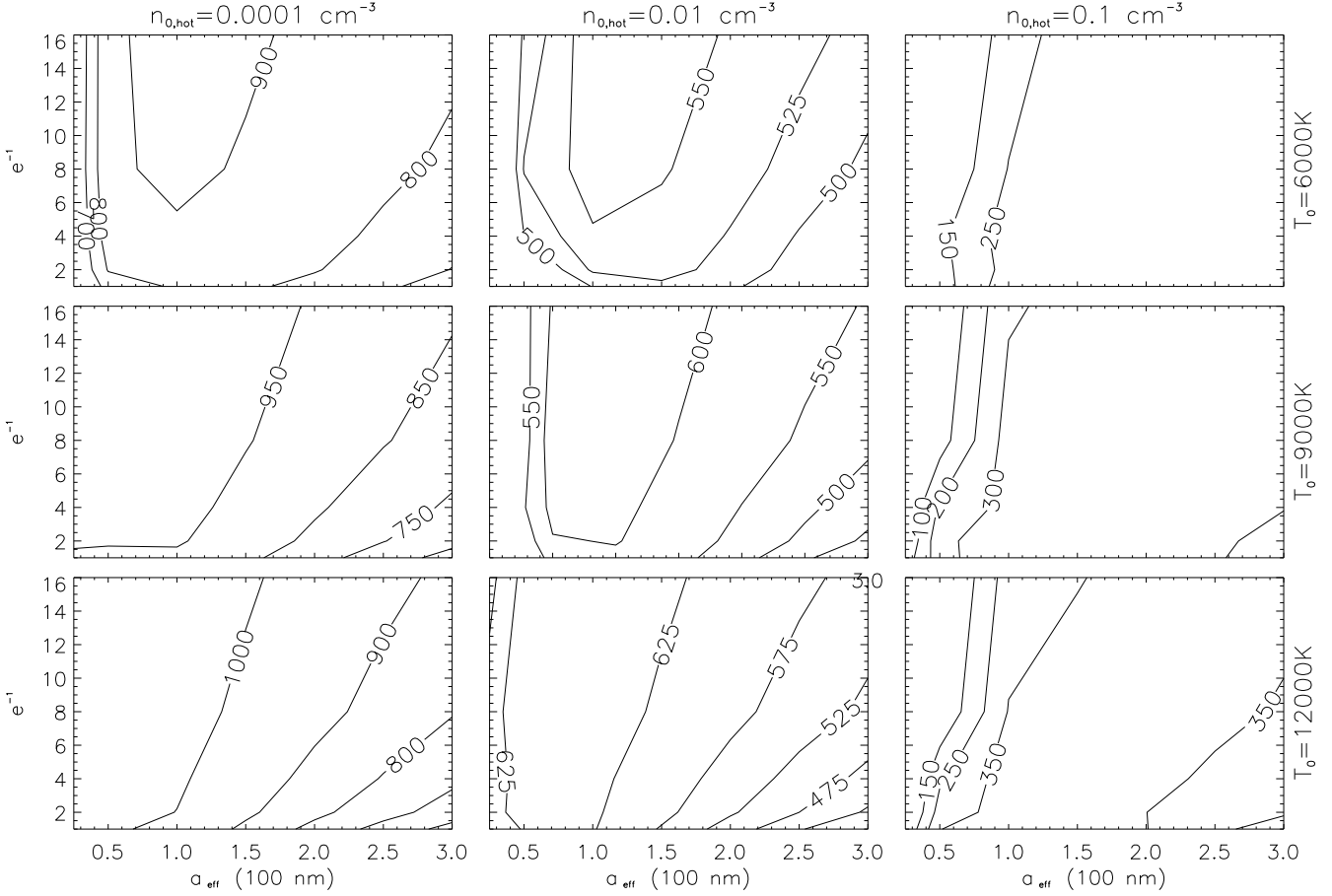}}
\caption{Contours of constant velocity in the $(a_\mathrm{eff}, 
e^{-1})$ plane for graphite like carbon grains (1000 C sample from
J\"ager {et al.}\ \cite{jager}).}
\label{gcontours}
\end{figure*}

Next, we have performed the full numerical solution to the dynamical equations
for ellipsoidal grains of different ellipticities and sizes.
Specifically we have calculated for 
$a_\mathrm{eff} = 25, 50, 100, 150, 200, 250, 
300\thinspace\mathrm{nm}$, and 
$e = 1, 1/2, 1/4, 1/8, 1/16$ for each $a_\mathrm{eff}$. The calculations
have been done for twelve different initial values of $R$, equally spaced from
1-12 kpc, and always starting at $z=0.2$ kpc.
The grains were followed for a period of 2 Gyr to determine their final
velocity. Figs.~\ref{tvr01} and~\ref{tvr1} show some examples of velocities
and displacements $r=(R^2+z^2)^{1/2}$ as a function of time, for the 
present day Milky Way and
for our extremely gas enriched model. In each case the velocity and 
displacement are shown for three combinations of $a_\mathrm{eff}$ and
$e$, namely $(a_\mathrm{eff},e)=(50\thinspace\mathrm{nm},1), 
(50\thinspace\mathrm{nm},1/8),(300\thinspace\mathrm{nm},1)$, and for two
different initial galactocentric distances $R_0=2,12\thinspace\mathrm{kpc}$.
We have used $Q$ values for the appropriate size and shape at
$T_0=9000\thinspace\mathrm{K}$. 
Both figures show that it is quite difficult to expel dust which has a 
large initial galactocentric distance, whereas most dust is expelled rapidly
from the galactic center. In the gas rich model, we find that grains
starting at $R=12$ kpc are almost completely confined by drag forces.
The figures also show that grains of different 
sizes and ellipticities behave fairly similarly.

After solving the dynamical equations, we have then calculated the final
velocity of the dust grains, averaged over the entire galaxy.
To do this we have assumed that the dust production rate is proportional to
the light profile of the disk, i.e.\ $R \propto e^{-R/R_d}$.
In some cases dust grains are confined within the galaxy by drag forces in the gas.
This happens particularly in the outer parts of the host galaxy and in the gas
rich model.  For such grains
we set the final velocity equal to zero. This means that our averaged velocity
is in fact an average over {\it all} grains, not only the ones which escape.
Thus, the averaged final velocities are in fact a folding of the velocity
of escaping dust grains and escape probabilities.

\begin{figure*}[t]
\resizebox{\hsize}{!}{\includegraphics{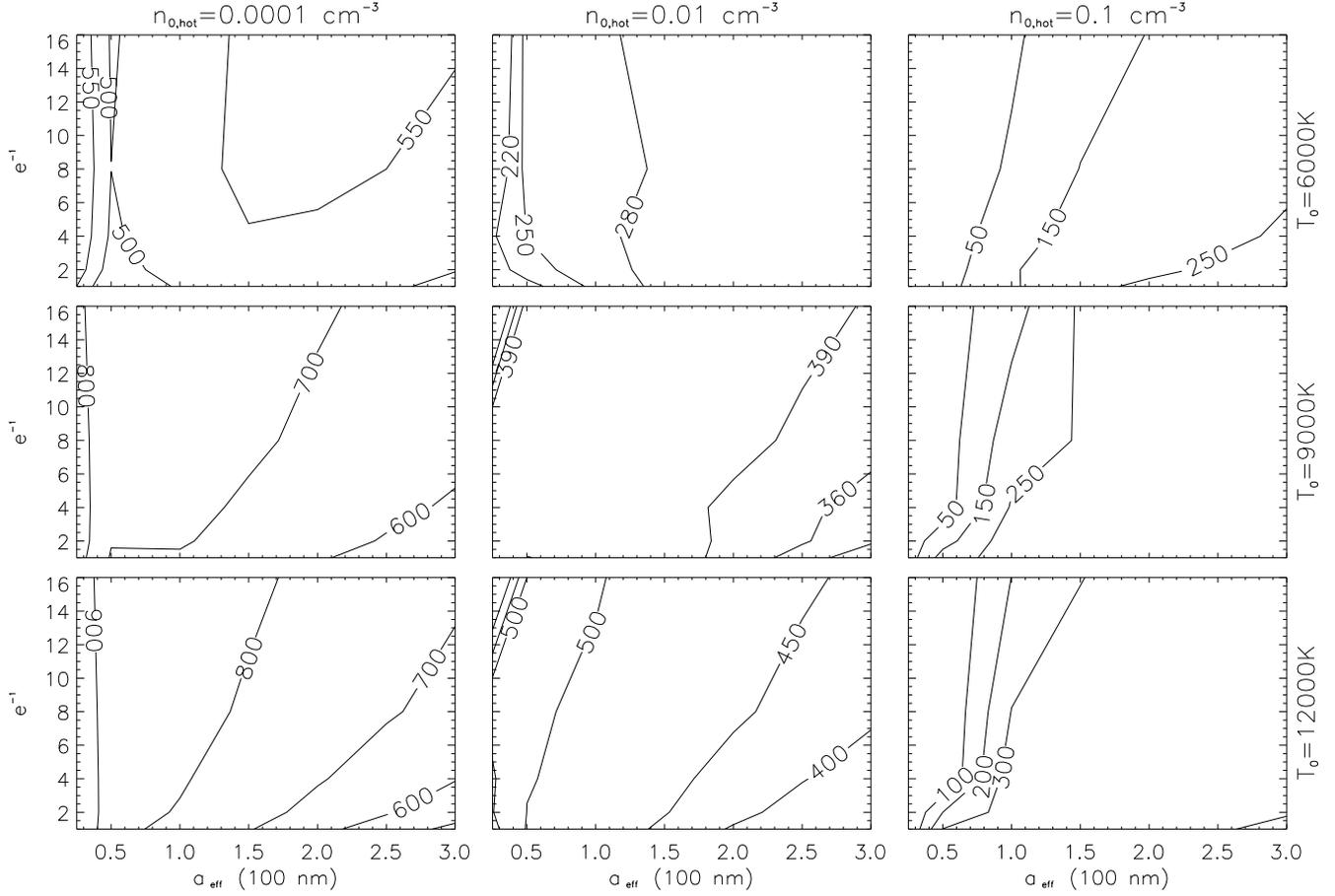}}
\caption{Contours of constant velocity in the $(a_\mathrm{eff}, 
e^{-1})$ plane for amorphous carbon dust grains
(400 C sample from
J\"ager {et al.}\ \cite{jager}).}
\label{ccontours}
\end{figure*}

Figs.~\ref{gcontours} and~\ref{ccontours} summarise our results for 
graphite and amorphous carbon respectively, showing contours of constant 
velocity in the $(a_\mathrm{eff}, e^{-1})$ plane, for all combinations of 
base density $n_0$ (represented by $n_{0,\mathrm{hot}}$ on top of the 
columns) and temperature $T_0$ (to the right of each row). 
Exactly as expected the averaged velocity is a decreasing function of the
gas density and an increasing function of emission temperature
(unless the grains are very large).

Our main result is that the final grain velocity is a very slowly varying
function of ellipticity. The reason is that although the effective $Q$-value
increases with ellipticity if the effective radius is kept fixed, the drag
force increases too. Drag therefore acts as a ``velocity-moderator'' in the
sense that it tends to erase velocity difference due to ellipticity.

The differences in final velocity due to varying grain size is straightforward
to understand. The acceleration due to radiation pressure is proportional
to $a_{\rm eff}^{-1}$ so that one should expect decreasing final velocities
for larger grains. However, the effective $Q$-value is a strongly increasing
function of $a_{\rm eff}$ for small grains (up to roughly 100 nm), so that
in the end there will be some $a_{\rm eff}$ where the velocity peaks.
Exactly this effect is seen in Figs.~3 and 4.

Note that in the upper left plot of fig.~\ref{ccontours} the velocity
increases again at very small radii. This is probably a numerical artifact
produced by uncertainty in the calculation of the effective $Q$-value
by the program DDSCAT. A few test runs have shown that increasing
the number of discrete dipoles in the grain does diminish the 
$Q$ value, but at much incresed computational time.

Our numerical results thus indicate that it will be difficult to 
separate dust grains of different ellipticities, but that in certain
cercumstances it could be possible to expel large grains only.

\subsection{Model dependencies}

We have investigated how dependent our conclusions are on our assumptions
about the drag force and the number density distribution of the gas, 
by performing some sample calculations using more conservative
assumptions.

\begin{figure*}[t]
\resizebox{\hsize}{!}{\includegraphics{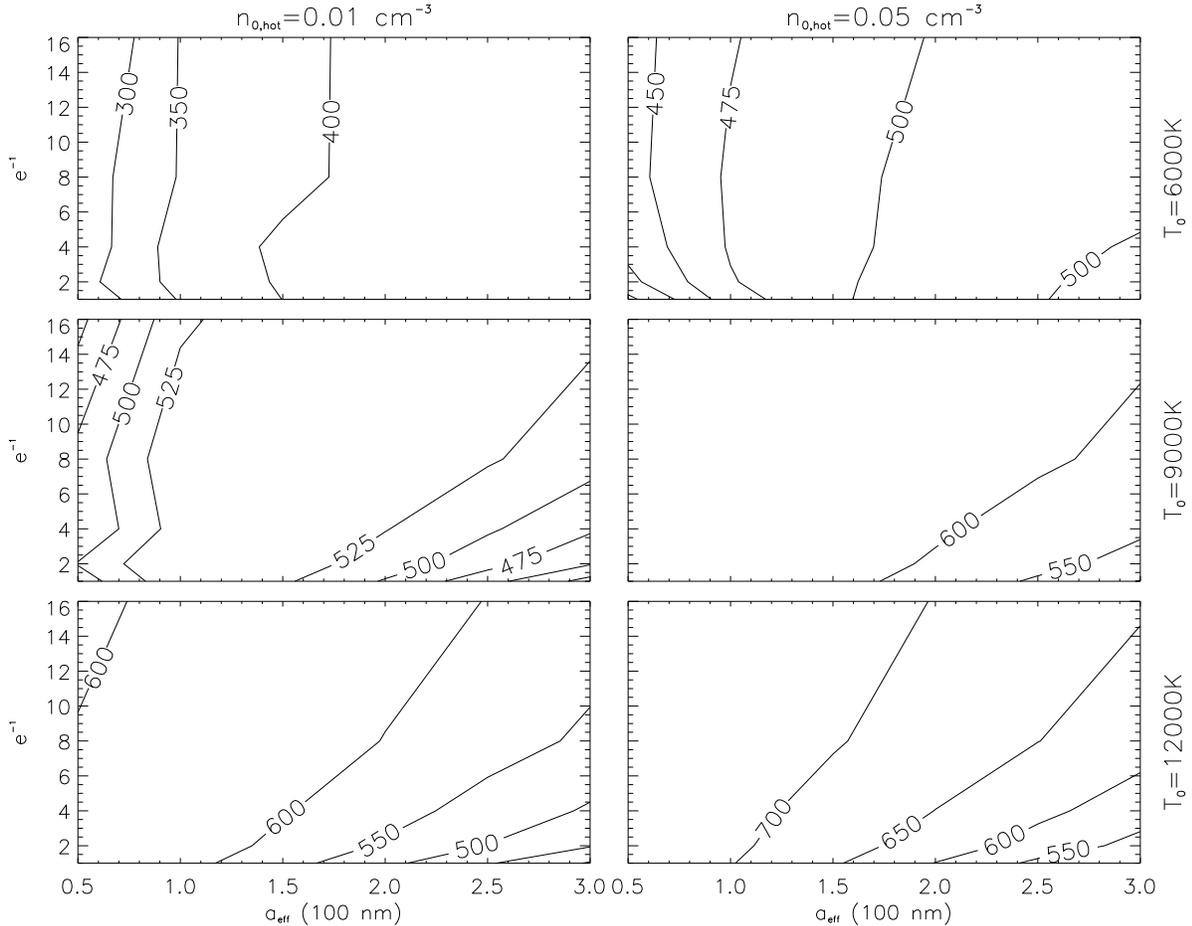}}
\caption{Contours of constant velocity in the $(a_\mathrm{eff}, 
e^{-1})$ plane for amorphous carbon dust grains
using the approximate drag force from Il'in (\cite{ilin}) (left column),
and with exponential fall-off in number density with $R$ (right column).}
\label{newcont}
\end{figure*}

In calculating the drag force, the grain velocities were
assumed to be large compared to the thermal velocities in the gas. But
adopting gas temperatures of $100\mathrm{K}$ and $10^6\mathrm{K}$ for the
cold and hot gas component respectively, the average thermal velocity
of the hot component is $v_\mathrm{gas} \simeq 130\mathrm{km/s}$, 
which is not negligible
compared to the smaller grain velocities. 
In order to check the validity of using our Eq.~(\ref{eq:drag}) we
recalculated the
final averaged velocities of amorphous carbon grains in the model 
with base densities
representative of the present day Milky Way (the second model), using the
approximate formula for the drag force given by Il'in (\cite{ilin}, Eq.\ 23). 
This approximation is designed to be valid for all grain velocities.
The results are shown in the first column of figure~\ref{newcont}.

Even though our equation can
produce errors of roughly 25\% in absolute velocities, it has no effect
on the conclusion that grain velocities are fairly independent of
size and ellipticity.

For the gas we assumed that the number density distribution was constant
as a function of $R$. We have investigated the effect of this by
performing a calculation using an exponential fall-off with both $R$ and
$z$. We use base densities of 
$n_{0,\mathrm{cold}}=1.1\thinspace\mathrm{cm}^{-3}$ and
$n_{0,\mathrm{hot}}=0.05\thinspace\mathrm{cm}^{-3}$, 
and the same scale length as the stellar disk.
This produces
densities equal to those of the present day Milky Way at
$R_0 \sim 8.5\thinspace\mathrm{kpc}$. Here we return to using
the expression for the drag force given by Eq.~\eqref{eq:drag}.
The results are shown
in the second column of figure~\ref{newcont}. 
Again, this yields results which are somewhat different in absolute
magnitude relative to those found from the constant density gas disk.
However, the velocity differences due to different size and ellipticity
are still small, which is the main point. 

In light of the large uncertanties in these calculations, because of our poor
knowledge of the nature of high-redshift galaxies, we think our approximations
are reasonable. They do yield different absolute results, but have no bearing
on our conclusion about the relative velocities of different dust grains.

\section{Discussion}

We have performed numerical calculations of how dust grains are expelled from
galaxies by the radiation pressure produced by starlight.
The calculations were done for a whole range of different sizes and shapes,
as well as for several different host galaxy types.
In all cases, final grain velocities are almost independent of ellipticity.
Our conclusion therefore is that the original scenario of Aguirre 
(\cite{aguirre}) does not work.

Also, if the galaxy does not contain significantly more gas
than the present day Milky Way, the relative velocity difference 
between grains of different sizes is quite small, unless the grains become
small.

The mechanism proposed by Aguirre (\cite{aguirre2}) requires that only dust 
with effective radius larger than about 100 nm should escape to 
the intergalactic medium,
in order not to have excessive reddening (this can be relaxed to about 50 nm
if only graphite grains are considered (Aguirre \cite{aguirre2})).
Our results indicate that this mechanism does not work for the two models
with the least gas content (the first two columns in 
Figs.~\ref{gcontours} and \ref{ccontours}), unless 
the smaller grains are destroyed completely by sputtering.
However, the findings of Ferrara {et al.} (\cite{ferrara}) 
and Davies {et al.} (\cite{davies}), indicate that sputtering is not 
effective unless the radius becomes as low as 10 nm.
On the other hand, Shustov \& Vibe (\cite{shustov}) find that grains up to 50 nm can be
destroyed.
In light of our poor understanding of the sputtering mechanism it is 
not possible to rule out the Aguirre model. However, just from looking
at grain velocities the model does seem an unlikely explanation.

As for the third model with extremely large gas content, there is a difference
between small and large grains, the smaller grains having lower velocities.
This seems to be exactly the type of selection effect we are looking for and,
furthermore, the model where it happens could be quite similar to high redshift
galaxies. However, the result should be treated with caution because 
this model is already close to the limit where all grains are completely
confined by gas drag. 
Altogether we find that there {\it are} models where dust segregation can be
produced, but that the physical conditions have to be just right.
We can therefore not rule out the Aguirre (\cite{aguirre2}) model, only say that
it requires some fine tuning.

Our results indicate that most of the dust produced in galaxies will
be expelled by radiation pressure. Of course we have neglected galactic sputtering
of grains, which might destroy very small grains. However, measurements indicate
that the intergalactic medium in fact has a fairly high metal content
(Mushotsky \cite{mushotsky}). 
This in turn
is likely to mean that most of the produced dust is expelled from the host
galaxies. However, it also seems very likely that almost all of the dust is
subsequently destroyed in the intracluster medium (Dwek et al. \cite{dwek},
Stickel et al. \cite{stickel}).
Note that since there is probably effective sputtering in the intracluster medium,
this could also mean that selection of large dust grains could be due to
sputtering in the ICM. This also limits our ability to rule out the 
Aguirre (\cite{aguirre2}) scenario. We can only say that sputtering within
galaxies is an unlikely explanation.

The above discussion does not take into account that radiation pressure 
may not even be a dominant mechanism for dust expulsion at high redshift. 
As mentioned above, Gnedin (\cite{gnedin}) has argued that galaxy mergers 
may provide the most efficient
mechanism for expulsion of grains into the intergalactic medium. If this is the
case there is even less reason for suspecting that grains with flat opacity
curves should preferentially be emitted, and the above scenario will be 
even less credible.

{\it Note added} -- After this paper had been submitted there has been
a growing interest in the possibility of progenitor evolution systematically
biasing the SNIa data. At present it seems unclear what the effects of evolution
are, but it has been pointed out that if there is evolution, then it will
be very difficult to extract cosmological parameters from the data
(Drell, Loredo \& Wasserman \cite{drell}; 
Dominguez et al. \cite{dominguez}; 
Riess et al. \cite{riess2};
Wang \cite{wang}).

\begin{acknowledgements}

We wish to thank Anja C.\ Andersen for many enlightening conversations 
regarding the properties of interstellar dust grains.
Also, we wish to thank the anonymous referee for many constructive and
enlightening comments on the initial version of this paper.

\end{acknowledgements}

\appendix
\section{Radiation pressure coefficients}
\label{qtables}

The following table shows the averaged radiation 
pressure coefficients $Q_\mathrm{pr}^*$ as a function of $a_\mathrm{eff}$,
$e$ and radiation field temperature $T_0$, calculated for amorphous
carbon and graphite. $a_\mathrm{eff}$ is in units of 100nm.

\begin{table}\centering
\caption{Effective radiation pressure coefficients, $Q_{\rm pr}^*$,
for different grain sizes and ellipticities.}
\begin{tabular}{l|lllll}
\hline\hline
\multicolumn{6}{c}{Carbon, $T_0=6000\thinspace\mathrm{K}$} \\
\hline
$a_\mathrm{eff}\backslash e$ &1&1/2&1/4&1/8&1/16 \\
\hline
0.25& 0.15 & 0.17 & 0.18 & 0.19 & 0.20 \\
0.5 & 0.33 & 0.36 & 0.38 & 0.39 & 0.40 \\
1.0 & 0.74 & 0.78 & 0.80 & 0.82 & 0.83 \\
1.5 & 1.15 & 1.19 & 1.21 & 1.24 & 1.25 \\
2.0 & 1.52 & 1.56 & 1.60 & 1.64 & 1.66 \\
2.5 & 1.82 & 1.88 & 1.94 & 2.01 & 2.06 \\
3.0 & 2.05 & 2.15 & 2.25 & 2.36 & 2.43 \\
\hline\hline
\multicolumn{6}{c}{Carbon, $T_0=9000\thinspace\mathrm{K}$} \\
\hline
0.25& 0.27 & 0.29 & 0.30 & 0.31 & 0.32 \\
0.5 & 0.56 & 0.59 & 0.61 & 0.63 & 0.63 \\
1.0 & 1.09 & 1.13 & 1.17 & 1.21 & 1.23 \\
1.5 & 1.52 & 1.58 & 1.64 & 1.71 & 1.78 \\
2.0 & 1.82 & 1.92 & 2.03 & 2.16 & 2.27 \\
2.5 & 2.01 & 2.18 & 2.34 & 2.54 & 2.70 \\
3.0 & 2.11 & 2.36 & 2.59 & 2.86 & 3.09 \\
\hline\hline
\multicolumn{6}{c}{Carbon, $T_0=12000\thinspace\mathrm{K}$} \\
\hline
0.25& 0.37 & 0.39 & 0.40 & 0.40 & 0.41 \\
0.5 & 0.72 & 0.75 & 0.77 & 0.79 & 0.80 \\
1.0 & 1.29 & 1.34 & 1.39 & 1.46 & 1.51 \\
1.5 & 1.67 & 1.77 & 1.87 & 2.01 & 2.12 \\
2.0 & 1.88 & 2.07 & 2.24 & 2.45 & 2.63 \\
2.5 & 1.99 & 2.26 & 2.50 & 2.81 & 3.07 \\
3.0 & 2.01 & 2.37 & 2.69 & 3.09 & 3.43 \\
\hline\hline
\multicolumn{6}{c}{Graphite, $T_0=6000\thinspace\mathrm{K}$.} \\
\hline
0.25& 0.26 & 0.31 & 0.35 & 0.39 & 0.40 \\
0.5 & 0.57 & 0.65 & 0.73 & 0.79 & 0.82 \\
1.0 & 1.26 & 1.38 & 1.48 & 1.56 & 1.61 \\
1.5 & 1.87 & 1.99 & 2.09 & 2.20 & 2.30 \\
2.0 & 2.30 & 2.45 & 2.57 & 2.73 & 2.89 \\
2.5 & 2.56 & 2.76 & 2.93 & 3.16 & 3.29 \\
3.0 & 2.69 & 2.95 & 3.18 & 3.51 & 3.82  \\
\hline\hline
\multicolumn{6}{c}{Graphite, $T_0=9000\thinspace\mathrm{K}$.}\\
\hline
0.25& 0.38 & 0.42 & 0.45 & 0.48 & 0.49 \\
0.5 & 0.79 & 0.86 & 0.92 & 0.96 & 0.98 \\
1.0 & 1.56 & 1.64 & 1.71 & 1.79 & 1.86 \\
1.5 & 2.07 & 2.20 & 2.31 & 2.45 & 2.59 \\
2.0 & 2.35 & 2.55 & 2.73 & 2.98 & 3.20 \\
2.5 & 2.47 & 2.76 & 3.02 & 3.37 & 3.70 \\
3.0 & 2.48 & 2.85 & 3.19 & 3.67 & 4.10   \\
\hline\hline
\multicolumn{6}{c}{Graphite, $T_0=12000\thinspace\mathrm{K}$.}\\
\hline
0.25& 0.45 & 0.49 & 0.51 & 0.53 & 0.54 \\
0.5 & 0.92 & 0.97 & 1.01 & 1.04 & 1.06 \\
1.0 & 1.69 & 1.75 & 1.81 & 1.90 & 1.98 \\
1.5 & 2.10 & 2.27 & 2.40 & 2.57 & 2.72 \\
2.0 & 2.28 & 2.54 & 2.78 & 3.10 & 3.34 \\
2.5 & 2.31 & 2.67 & 2.99 & 3.45 & 3.85 \\
3.0 & 2.25 & 2.70 & 3.12 & 3.68 & 4.22 \\
\hline
\end{tabular}
\end{table}


\begin{thebibliography}{}

\bibitem[1999a]{aguirre}
Aguirre, A.N., 1999a, Astrophys.\ J.\ Lett.\, 512, L19.

\bibitem[1999b]{aguirre2}
Aguirre, A.N., 1999b, astro-ph/9904319.

\bibitem[1989]{barsella}
Barsella, B. et al., 1989, A\&A, 209, 349.

\bibitem[1987]{binney}
Binney, J. \& Tremaine, S., 1987, Galactic Dynamics, J.P. Ostriker (ed.), Princeton.

\bibitem[1989]{ccm89}
Cardelli, J.A., Clayton, C.G., and Mathis, J.S., 1989, 
Astrophys.\ J.\, 345, 245.

\bibitem[1998]{davies}
Davies, J.I. et al., 1998, MNRAS, 300, 1006.

\bibitem[1999]{dominguez}
Dominguez, I. et al., 1999, astro-ph/9905047.

\bibitem[1998]{draine}
Draine, B.T. \& Flatau, P.J., 1998, Princeton observatory preprint POPe-785.     

\bibitem[1984]{drlee}
Draine, B.T. \& Lee, H., 1984, Astrophys.\ J., 285, 89. 

\bibitem[1979]{ds79}
Draine, B.T. \& Salpeter, E.E., 1979, Astrophys.\ J., 231, 77. 

\bibitem[1999]{drell}
Drell, P.S., Loredo, T.J. \& Wasserman, I., 1999, astro-ph/9904111.

\bibitem[1990]{dwek}
Dwek, E., Rephaeli, Y. \& Mather, J.C., 1990, Astrophys.\ J.\, 350, 104.

\bibitem[1991]{ferrara}
Ferrare, A. et al., 1991, Astrophys.\ J., 381, 137.

\bibitem[1998]{fukuda}
Fukuda, Y. et al., 1998, Phys.\ Rev.\ Lett.\, 81, 1562.

\bibitem[1998]{garnavich}
Garnavich, P.M. et al., 1998,
astro-ph/9806396 (to appear in Astrophys.\ J.).

\bibitem[1998]{gnedin}
Gnedin, N.Y., 1998, MNRAS, 294, 407.

\bibitem[1991]{heiles}
Heiles, C., 1991, {\it The Interstellar Disk-Halo Connection 
in Galaxies}, proceedings of the 144th symposium of the IAU,
Leiden 1990, Kluwer Academic Publishers.

\bibitem[1998]{hwt98}
H{\"o}flich, P., Wheeler, J.C., and Thielemann, F.K., 1998,
Astrophys.\ J., 495, 617 (1998).

\bibitem[1994]{ilin}
Il'in, V.B., 1994, A\&A, 281, 486.

\bibitem[1998]{jager}
J\"ager, C., Mutschke, H. \& Henning, T., 1998, A\&A, 332, 291-299.

\bibitem[1990]{kt90}
Kolb, E.W. \& Turner, M.S., 1990, {\it The Early Universe},
Addison-Wesley Press.

\bibitem[1991]{lockman}
Lockman, F.J., 1991, {\it The Interstellar Disk-Halo Connection 
in Galaxies}, proceedings of the 144th symposium of the IAU,
Leiden 1990, Kluwer Academic Publishers.

\bibitem[1996]{mushotsky}
Mushotsky, R. et al., 1996, Astrophys.\ J.\, 466, 686.


\bibitem[1993]{peebles}
Peebles, P.J.E., 1993, {\it Principles of Physical
Cosmology}, Princeton University Press.

\bibitem[1997]{perlmutter97}
Perlmutter, S. et al., 1997, Astrophys.\ J.\, 483, 565.

\bibitem[1998]{perlmutter98}
Perlmutter, S. et al., 1998,
astro-ph/9812133 (to appear in Astrophys.\ J.).

\bibitem[1998]{riess}
Riess, A.G. et al., 1998,
astro-ph/9805201 (to appear in Astron. J.).

\bibitem[1999]{riess2}
Riess, A.~G. et al., 1999, astro-ph/9907038.

\bibitem[1998]{schmidt}
Schmidt, B.P. et al., 1998,
astro-ph/9805200 (to appear in Astrophys.\ J.).

\bibitem[1995]{shustov}
Shustov, B.M. \& Vibe, D.Z., 1995, Astron.\ Zhurnal, 72, 650.

\bibitem[1998]{stickel}
Stickel, M. et al., 1998, A\&A, 329, 55.

\bibitem[1994]{suchkov}
Suchkov, A. et al., 1994, Astrophys.\ J.\, 430, 511.

\bibitem[1999]{wang}
Wang, Y., 1999, astro-ph/9907405.

\end{thebibliography}
\end{document}